\documentclass[pra,
                twocolumn,
                preprintnumbers,
                amsmath,
                amssymb,
                showpacs,
                superscriptaddress]{revtex4-2}
\usepackage{graphicx} 
\usepackage{dcolumn}  
\usepackage{bm}       
\usepackage{amsfonts}
\usepackage{mathtools}
\usepackage[colorlinks=true,linkcolor=black,citecolor=red,urlcolor=blue]{hyperref}
\usepackage{color}
\usepackage{natbib}
\usepackage{float}
\usepackage{scalerel}
\usepackage{braket}

\newcommand{\ful}{\mbox{C$_{\mbox{\scriptsize{240}}}$}}
\newcommand{\na}{\mbox{Na$_{\mbox{\scriptsize{20}}}$}}
\newcommand{\naful}{\mbox{Na$_{\mbox{\scriptsize{20}}}$}@\mbox{C$_{\mbox{\scriptsize{240}}}$}}

\newcommand{\eq}[1]{Eq.~(\ref{#1})}

\begin{document}

\title{Hybridization and coherence in subshell differential intercluster plasmonic decay in $\naful$}

\author{Rasheed Shaik}
\affiliation{School of Physical Sciences, Indian Institute of Technology Mandi, Kamand, H.P. 175075, India}

\author{Hari R. Varma}
\email[]{hari@iitmandi.ac.in}
\affiliation{School of Physical Sciences, Indian Institute of Technology Mandi, Kamand, H.P. 175075, India}

\author{Mohamed El-Amine Madjet}
\affiliation{Department of Natural Sciences, D.L.\ Hubbard Center for Innovation and Entrepreneurship, Northwest Missouri State University, Maryville, Missouri 64468, USA}
\affiliation{Bremen Center for Computational Materials Science, University of Bremen, Bremen 28359, Germany}

\author{Fulu Zheng}
\affiliation{Department of Physics, School of Physical Science and Technology, Ningbo University, Ningbo, 315211, P.R. China}
\affiliation{Bremen Center for Computational Materials Science, University of Bremen, Bremen 28359, Germany}

\author{Himadri S. Chakraborty}
\email[]{himadri@nwmissouri.edu}
\affiliation{Department of Natural Sciences, D.L.\ Hubbard Center for Innovation and Entrepreneurship, Northwest Missouri State University, Maryville, Missouri 64468, USA}

\date{\today}



\begin{abstract}
We study the ground state structure and aspects of photoionization dynamics of the $\naful$ endofullerene. The structure shows effects from the electronic coupling between the nested cluster and the \textcolor{black}{fullerene} cage. They include the (i) alterations of the overall potential, and thus, the force field, (ii) electron transfer from the cluster to the fullerene forming ionic units, and (iii) hybridization from the admixture of free $\na$ occupied levels with \textcolor{black}{experimentally known} super-atom molecular orbital (SAMO) type empty levels of $\ful$ \textcolor{black}{accessible in the jellium-DFT model}. These modifications influence the photoionization dynamics of the endofullerene. For the high energy ionization of $\na$-type levels, a significant overall enhancement of the cross section is noted \textcolor{black}{from additional ionizing force that $\ful$ offers}. More remarkably, the photoexcited plasmons, \textcolor{black}{both the giant plasmon and the higher energy plasmon,} in $\ful$ decay in parts through $\na$ ionization continuum via the resonant intercluster Coulombic decay (ICD) process. These lead to dramatic enhancements in the ionization of individual $\na$-type levels \textcolor{black}{resulting enhancements in the cluster's total ionization yield}. Based on hybridization, this enhancement incorporates a coherent mixing of the ICD and SAMO-induced Auger-decay amplitude, in which the ICD contribution is dominant. 
\end{abstract}

\maketitle

\section{Introduction}

With their unique feature of an empty interior, fullerenes inspire novel research ideas by enclosing atoms, molecules, smaller fullerenes, or clusters within their carbon network. Resulting endofullerene complexes, denoted as A@C$_{n}$ (A being the encapsulated species and C$_{n}$ the fullerene cage), offer molecular laboratories to probe mutual interactions. Synthesis methods of such nested systems, having characteristically weak A-C$_{n}$ bonding, are rapidly evolving~\cite{Popov2017} which piggyback the advantage of their higher stability at room temperature. This arena of activities not only \textcolor{black}{motivates fundamental research} but also holds prospects of applications. The use of endofullerenes as qubits for quantum computing, by harnessing the electron spin of the encapsulated dopant, has been discussed~\cite{Harneit2017}. The materials also find applications in the areas of, for example, superconductivity~\cite{Iwasa2010}, energy storage~\cite{Jiang2021}, and drug delivery~\cite{Dellinger2013}.

An intriguing aspect of the study of endohedral systems involves the investigation of the effect of fullerene cage on the spectroscopy of the encapsulated species, and vice versa. In particular, the photoelectron spectra of these composites are significantly richer compared to the spectra of independent systems~\cite{Deshmukh2021}. A theoretical study by Averbukh and Cederbaum~\cite{Averbukh2006} first revealed a variety of intersite Coulombic decay (ICD) pathways within an endohedral fullerene complex, using the Ne@C$_{60}$ as a prototype. Later, in a series of studies involving C$_{60}$ with varieties of confined atoms a host of ICD processes~\cite{javani2014,magrakvelidze2016,de2016,khokhlova2020,de2021} for single-electron vacancy decay resonances was predicted. ICD is a process in which a weakly bounded complex of multiple units may absorb extreme ultraviolet (XUV) to x-ray range photons inducing inner-shell excitations in one unit which result in the outer-shell ionization at a neighboring unit. This non-local decay mechanism is facilitated by the Coulomb interaction between the sites. Obviously, isolated units that make up the complex cannot include these decay processes on their own. In general, ICD processes leading to non-radiative decays have been well researched~\cite{Jahnke2020} since its original discovery by Cederbaum \textit{et al.}~\cite{Cederbaum1997} more than two decades ago.   

Traditional methods of electron~\cite{Marburger2003,Ren2016} and ion~\cite{Wiegandt2019} spectroscopy using various coincidence techniques~\cite{LaForge2016} are employed to probe ICD signatures. Present-day pump-probe approaches~\cite{Schnorr2013,Takanashi2017}, particularly those involving light field streaking techniques~\cite{Trinter2013}, have provided access to time-resolved ICD dynamics. A recent comprehensive review of ICD research carried out till 2020 on a range of materials and potential applications can be found in Ref.\,\cite{Jahnke2020}. More recently, interesting new scenarios are added to the ICD landscape~\cite{Cederbaum2021,Barik2022}. The signature of ICD pathway between holmium nitride and the surrounding fullerene cage was detected by employing the ion-electron coincidence spectroscopy combined with velocity map imaging (VMI) technique~\cite{Obaid20}. The first measurement of ICD in liquid water has been reported~\cite{Zhang2022} which underscores the relevance of studying ICD in the context of radiation damage in aqueous environment. The role of ICD in the formation of complex molecules in the interstellar medium has of late been uncovered~\cite{Barik23}. A very recent experimental investigation using high resolution electron spectroscopy and VMI techniques has unraveled efficient resonant ICD (RICD) signals in He nanodroplets irradiated by weak synchrotron radition~\cite{SRK24}.

The Na cluster-enclosing endofullerenes~\cite{trujillo1996,yang2017} provide a unique platform to study ICD phenomena between two clusters involving the decay of collective resonances. The characteristic existence of the giant plasmon resonance \textcolor{black}{(GPR)} \textcolor{black}{and a higher energy plasmon resonance (HPR)} in C$_{n}$ which can be excited by the XUV radiation makes the cage environment very special.  During the de-excitation of these plasmons, it is possible that some of their released energy is resonantly transferred to the ionization channels of the caged cluster. This may facilitate efficient RICD pathways of the plasmon decay within the ionization dynamics. Indeed, in our recent photoionization study of $\naful$~\cite{shaik2023}, we have predicted such plasmonic RICD \textcolor{black}{of GPR} leading to significant oscillator strength transfer in the XUV spectrum.  The strength of this ICD plasmon is found comparable with the native plasmon resonance that occurs in the absorption spectrum of free $\na$. \textcolor{black}{But, RICD of HPR is yet to be probed.} 

In the current study, we first scrutinize the facets of modification of the ground state structure of $\naful$ arisen from the electronic coupling between $\na$ and $\ful$. To do so, we first used a quantum chemical (QC) calculation for bench-marking the details of the structure. We then proceed to examine how these facets modify the properties of photoionization. We particularly focus on the plasmonic RICD \textcolor{black}{of both GPR and HPR} that plays out for individual $\na$-type subshell emission. We further demonstrate an interpretation of the mechanism within the framework of a coherence condition between a dominant ICD channel and an {\em induced} Auger decay channel.

The paper is organized as follows: In section II, a description of the theory is presented. Section III-A discusses the results of ground state structure of $\naful$, while Section III-B includes the photoionization results within the limit of an independent particle calculation. Section III-C presents the plasmonic RICD in the photoionization cross section in a many-body framework and their interpretation to draw insights. We finally conclude our study.

\section{Theory}

\subsection{Quantum chemical ground state}
To optimize the ground state geometry of $\na$ and $\naful$, a QC methodology within the density functional theory (DFT) is first employed using the Turbomole software~\cite{turbomole}. The computational approach utilizes the B3LYP exchange-correlation (xc) functional~\cite{b3lyp1,b3lyp2} with the DFT-D3 dispersion correction~\cite{dft-d3}. We deploy an all-electron def2-SV(P) basis set~\cite{def2-sv} for these calculations. \textcolor{black}{The convergence criterion for self-consistent field energy and total energy is 10$^{-7}$ Hartree. In the geometry optimization step, the DFT-D3 dispersion correction is considered as well and the maximum norm of Cartesian gradient is converged to 10$^{-3}$ a.u. In the current work, we do not explore the effect of considering different $\na$ isomers. It may also be likely that $\na$ being caged in $\ful$ will restrict the possibility for the cluster to explore isomers.}

The detailed information obtained from the QC results forms the foundation in determining the system's ground state within a jellium-based DFT frame, aiming for the best possible match. This transition to jellium-DFT (jDFT) was necessary, since the description of the electron continuum \textcolor{black}{in a many-body frame, required to treat the plasmonic ionization process, is computationally very challenging} within a QC model. \textcolor{black}{We emphasize that plasmons being of a collective-electron origin are inaccessible in a single-electron mean-field theory, even in QC and even if it can handle single-electron resonances.}

\subsection{Jellium-DFT ground state}
Based on the detailed description in Ref.~\cite{madjet2008photoionization,choi2017}, \textcolor{black}{and following the Supplementary Material (SM) in Ref.~\cite{shaik2023}}, we model the spherically symmetric ground state structure of $\naful$ in a jDFT framework. The employed jellium representation of the ionic core has past success for systems with delocalized electrons, particularly in extending the method to include collective effects. In this, the atomistic cores of 20 Na$^{+}$ and 240 C$^{4+}$ ions of Na$_{20}$ and C$_{240}$ are replaced by classical electrostatic potentials obtained upon uniformly smudging the charges into, respectively, a sphere concentric within a shell. The potential $V_{\mbox{jel}}(\mathbf{r})$ thus generated radially distributes as:
\[V_{\mbox{\scriptsize jel}}(r)=\begin{cases}
              -\frac{Z_{c}}{2R_{c}}(3-(\frac{r}{R_{c}})^{2})-\frac{3Z_{f}}{2}\frac{(R_{0}^{2}-R_{i}^{2})}{R_{0}^{3}-R_{i}^{3}}, r\leq R_{c}\\\\
              -\frac{Z_{c}}{r}-\frac{3Z_{f}}{2}\frac{(R_{0}^{2}-R_{i}^{2})}{R_{0}^{3}-R_{i}^{3}}, R_{c}<r<R_{i}\\\\
              -\frac{Z_{c}}{r}-\frac{Z_{f}}{2}\frac{(3R_{0}^{2}-r^{2}-2R_{i}^{3}/r)}{R_{0}^{3}-R_{i}^{3}}, R_{i}\leq r \leq R_{0}\\\\
              -\frac{Z_{c}}{r}- \frac{Z_{f}}{r}, r>R_{0}
              \end{cases}.
\]
Here $Z_{c}=20$ and $Z_{f}=960$ are, respectively, the charges of the jelliumized core of Na$_{20}$ and C$_{240}$. The radius of C$_{240}$ is taken to be $R$ = 13.5 a.u.~\cite{choi2017} and that $R_{c}$ of Na$_{20}$ is calculated to be 10.7 a.u.~\cite{shaik2021}. $R_{i} = R-\Delta/2$ and $R_{0} = R+\Delta/2$ are the inner and outer radius of the $\ful$ shell which has a width ($\Delta$) of 2.87 a.u.\ as was determined in Ref.\,\cite{choi2017}. We used a parametrized pseudopotential, $V_{ps}(r)$ as,  
 \[V_{\mbox{\scriptsize ps}}(r)=\begin{cases}
 
              \frac{V_{0c}}{1+\exp{(r-R_{a})/\gamma}}-\frac{V_{0f}}{1+\exp{(R_{a}-r)/\gamma}}, 0 \leq r \leq R_{0}\\\\
              
              0, r>R_{0}
              \end{cases}.
\]
\textcolor{black}{The values of $V_{0c}$ and $V_{0f}$ were earlier chosen as constant pseudopotentials to model independent $\na$~\cite{shaik2021} and $\ful$~\cite{choi2017} that best reproduced their ionization potentials. However, in the above equation, they were needed to be further adjusted in order to imitate the QC electronic configuration. This gave $V_{0c}$ = 0.0057 a.u.\ and $V_{0f}$ = 0.45 a.u., respectively, in Na$_{20}$ and C$_{240}$ radial regions. Further, $R_{a}$ = $\frac{R_{c}+R_{i}}{2}$ = 11.36 a.u., $\gamma$ = 0.3 a.u., and the functional form is used to smoothly match between the two regions. We note, the adjustable pseudopotential~\cite{puska1993} in general provides a simplified description of the effects of the ion core, extending beyond the jellium model.} The Kohn-Sham equations for a system of 980 electrons, comprised of four valence electrons (2$s^{2}$2$p^{2}$) from each of 240 C atoms and the solo valence electron (3$s^{1}$) from each of 20 Na atoms, are then solved in the following manner.

The ground state self-consistent field potential \textcolor{black}{in the local density approximation (LDA)} can be written as,
\begin{equation}\label{dft-pot}
V^{\scriptsize \mbox{LDA}}(\mathbf{r}) = V_{\mbox{\scriptsize jel}}(\mathbf{r}) + \int d\mathbf{r}'\frac{\rho(\mathbf{r}')}{|\mathbf{r}-\mathbf{r}'|} + V_{\scriptsize \mbox{xc}}^{\scriptsize \mbox{LDA}}[\rho(\mathbf{r})],
\end{equation}
where $\rho(\mathbf{r})$ denotes the single-electron density. The second and the third term on the right-hand-side of \eq{dft-pot} are the direct and the xc contributions respectively. Following ~\cite{gunnarsson1976}, $V_{\scriptsize \mbox{xc}}^{\scriptsize \mbox{LDA}}$ can use a \textcolor{black}{basic LDA parametric form due to Gunnarsson and Lundqvist (GL)} in terms of $\rho(\mathbf{r})$:
\begin{eqnarray}\label{gl}
V_{\scriptsize \mbox{xc}}^{\scriptsize \mbox{LDA}}[\rho(\mathbf{r})] & = & -\left(\frac{3\rho(\mathbf{r})}{\pi}\right)^{1/3} \nonumber \\
 &-& 0.0333\log\left[1 + 11.4\left(\frac{4\pi\rho(\mathbf{r})}{3}\right)^{1/3}\right].\nonumber \\ 
\end{eqnarray}
Using a variational approach, the first term on the right-hand-side of \eq{gl} is exactly derivable from the Hartree-Fock (HF) exchange energy of a homogeneous electron gas with a uniform positively charged background~\cite{gunnarsson1976}. The second term is the conventional correlation potential not included in HF formalism~\cite{gunnarsson1976}.

To eradicate the unrealistic self-interaction of the $i^{th}$ occupied subshell, a correction scheme is employed from the outset~\cite{choi2017,perdew1981}:   
\begin{eqnarray}\label{dft-pot-sic}
 V^{i}(\mathbf{r}) &  = & V_{\mbox{\scriptsize jel}}(\mathbf{r}) + \int d\mathbf{r}'\frac{\rho(\mathbf{r}')-\rho_{i}(\mathbf{r}')}{|\mathbf{r}-\mathbf{r}'|} +  V_{\scriptsize \mbox{xc}}^{\scriptsize \mbox{LDA}}[\rho(\mathbf{r})] \nonumber \\ & - & V_{\scriptsize \mbox{xc}}^{\scriptsize \mbox{LDA}}[\rho_{i}(\mathbf{r})].
\end{eqnarray}
\textcolor{black}{The correction for the xc potential adopts the approach used routinely~\cite{pacheco1992,jena1992}}. The scheme obviously renders the potentials in \eq{dft-pot-sic} orbital-specific. Even though the approach entails an artificial correction for self-interaction, it is known to lead to a significant improvement in the asymptotic behavior of the potential and can produce ground state results comparable to what can be found with more contemporary xc functionals for free Na$_{20}$~\cite{shaik2021} and empty C$_{240}$~\cite{choi2017}. Therefore, results based on this approach are also used in the current study for $\na$ and $\ful$. \textcolor{black}{Furthermore, and rather crucially, the approach provided the best comparison of jDFT versus QC ground state of $\naful$ as will be seen in subsection III\,A.}

\subsection{DFT photoionization}
The photoionization dipole-response of a system to the external electromagnetic field is accounted in a linear-response time-dependent DFT (LR-TDDFT) technique\textcolor{black}{~\cite{madjet2008photoionization,choi2017}; alse see SM of Ref.\,\cite{shaik2023}. Besides fundamental predictions in C$_{60}$-based ICD studies~\cite{javani2014,magrakvelidze2016,de2016,de2021}, this approach in general has some good success in experiment-theory joint campaigns~\cite{ruedel2002}, including on plasmonic dynamics~\cite{scully2005,biswas2022}.} Here, the linear polarization of the field is set along the $z$ axis. LR-TDDFT perturbatively induces a frequency-dependent change in the electron density~\cite{grosseku1996},
\begin{equation}\label{ind-dens}
\delta \rho (\mathbf{r}^{\prime}; \omega) = \int \chi (\mathbf{r}, \mathbf{r}^{\prime}; \omega)
z  d\mathbf{r},
\end{equation}
which is a complex quantity. $\chi$ is the full susceptibility of the system which includes important many-body electron correlations. \eq{ind-dens} can be recast showing the independent particle (IP) susceptibility, $\chi^{0}$, and the complex perturbation potential, $\delta V$, as,
\begin{equation}\label{ind-dens-ip}
\delta \rho (\mathbf{r}^{\prime}; \omega) = \int \chi^{0} (\mathbf{r}, \mathbf{r}^{\prime}; \omega)
\delta V(\mathbf{r}; \omega) d\mathbf{r}.
\end{equation} 
\textcolor{black}{Ground-state {\em occupied} single-electron orbitals $\phi _{nl}$ and energies $\epsilon _{nl}$ can construct $\chi^{0}$~\cite{feibelman1975} in a Green's function formalism as follows:}
\begin{eqnarray}\label{suscep}
\chi^{0} (\mathbf{r},\mathbf{r}^{\prime };\omega) &=&\sum_{nl}^{occ}\phi _{nl}^{*}
(\mathbf{r})\phi _{nl}(\mathbf{r}^{\prime })\ G(\mathbf{r},\mathbf{r}^{\prime };\epsilon
_{nl}+\omega)  \nonumber \\
&+&\sum_{nl}^{occ}\phi _{nl}(\mathbf{r})\phi _{nl}^{*}(\mathbf{r}^{\prime })\ G^*
(\mathbf{r},\mathbf{r}^{\prime };\epsilon _{nl}-\omega).  
\end{eqnarray}

The Green's function with a parameter $E$ for a spherical system can be expanded in a spherical basis set to write as, 
\begin{equation}\label{eq1}
G (\mathbf{r},\mathbf{r'};E) =\sum_{lm}G_{lm}(r,r';E)Y_{lm}^{*}(\Omega)Y_{lm}(\Omega'),
\end{equation}
where the radial component $G_{lm}(r,r';E)$ must solve the radial equation:
\begin{eqnarray}\label{eq3}
\left(\frac{1}{r^{2}}\frac{\partial}{\partial r}r^{2}\frac{\partial}{\partial r}-\frac{l(l+1)}{r^{2}}-V+E\right)G_{lm}(r,r';E)\nonumber\\=\frac{\delta(r-r')}{r^{2}}.
\end{eqnarray}
\textcolor{black}{The continuum wave functions are calculated in the state dependent SIC potentials, therefore $G_{lm}(r,r';E)$ should be also state dependent, but for simplicity it is not shown.} Obviously, the regular $f_{l}(r;E)$ and the irregular $g_{l}(r;E)$ solution of the homogeneous partner of \eq{eq3} construct $G_{lm}$ to obtain,
\begin{equation}\label{green}
G_{lm}(r,r';E)=\frac{2f_{l}(r_{<};E)h_{l}(r_{>};E)}{W [f_{l},h_{l}]}.  
\end{equation}
Here $W$ is the Wronskian and $h_{l} = g_{l} + i f_{l}$.
$\delta V$ can then be explicitly written by the relation:
\begin{equation}\label{tot-pot}
\delta V(\mathbf{r};\omega)=z+V_{\mbox{ind}}(\mathbf{r};\omega),
\end{equation} 
where the complex induced potential
\begin{eqnarray}\label{ind-pot}
V_{\mbox{ind}}(\mathbf{r};\omega)=\int\frac{\delta \rho(\mathbf{r'};\omega)}{|\mathbf{r}-\mathbf{r}'|}d \mathbf{r}'+\left[\frac{\partial V_{xc}^{\scriptsize \mbox{LDA}}}{\partial \rho}\right]_{\rho=\rho_{0}}\delta \rho(\mathbf{r};\omega) \nonumber
\end{eqnarray} 
thus embodies the correlations, beyond the ground state DFT in \eq{gl}, and accounts for various many-body effects, including the collective plasmon resonances.

At the same time, $\chi$ and $\chi^{0}$ are related via the matrix equation,
\begin{equation}\label{rpa}
\chi = \chi^{0}\Bigg[1-\frac{\partial V}{\partial \rho}\chi^{0}\Bigg]^{-1}
\end{equation} 
\textcolor{black}{Due the state dependence of the ground state potential in the SIC method, the induced potential, the response function and the residual interaction should also be state dependent. But for simplicity following the approach by Saito et al.\ expressed in Eq.\,(5) of Ref.~\cite{saito1991}, we use
\begin{equation}
\frac{\partial V}{\partial \rho} = \frac{(N-1)}{N}\frac{\partial V}{\partial \rho}^{\scriptsize \mbox{LDA}},
\end{equation}
}
based on the variation of $V$ with respect to $\rho$. \textcolor{black}{Here, $N$ is the number of electrons.} \eq{rpa} can be solved for $\chi$ by matrix inversion method~\cite{bert1990} which in turn is used for evaluating $\delta \rho$ and thereby $\delta V$ by applying \eq{ind-dens} and \eq{tot-pot} self-consistently.

Finally, the photoionization (PI) cross section corresponding to a bound-to-continuum dipole transition is computed as the linear sum of independent subshell cross sections, $\sigma_{nl \rightarrow kl'}$, to read as:
 \begin{equation}\label{pi}
 \sigma_{\mbox{\scriptsize PI}}(\omega)=\sum_{nl}\sigma_{nl \rightarrow kl'}\sim\sum_{nl}2(2l+1)|\bra{kl'}\delta V\ket{nl}|^{2}.
\end{equation}  
Obviously, in \eq{pi}, we replace $\delta V$ by $z$ (in the radial-gauge) to calculate the uncorrelated IP result in LR-DFT.

The radial component $R_{k l'}$ of the computed final-state continuum wave function $\psi_{k l'}$ has the appropriate asymptotic behavior:
\begin{eqnarray}\label{rad_asymp}
\lim_{r\rightarrow\infty} R_{k l'} (r) &\sim&  \lim_{r\rightarrow\infty} [\cos (\delta_{l'}) f_{l'} (kr)
                 + \sin (\delta_{l'})g_{l'} (kr)] \nonumber \\
                 &=& \sin \left( kr - \frac{1}{2}\pi + \frac{z}{k}\ln (2kr) + \zeta_{l'} + \delta_{l'}\right)\nonumber \\
\end{eqnarray}
where $\zeta$ and $\delta$ are respectively the Coulomb and short-range phase-shifts seen by the photoelectron.

\begin{figure*}[ht!]
\includegraphics[width=16cm]{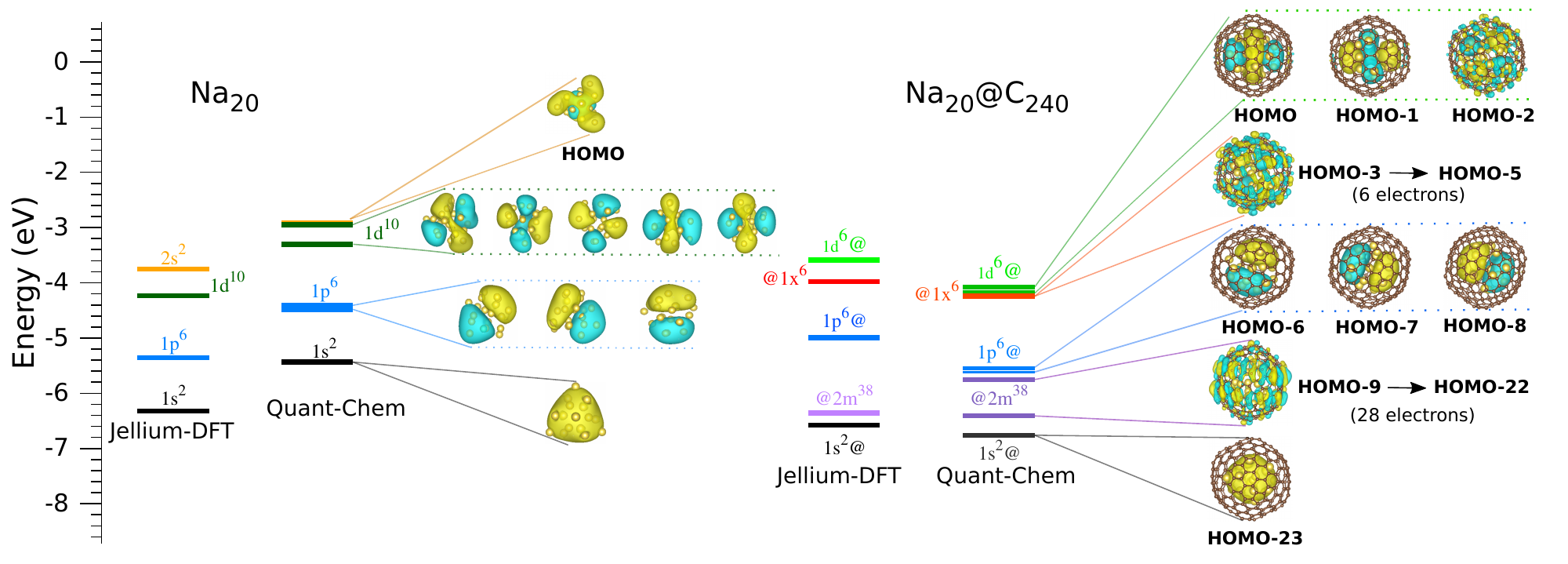}
\caption{All occupied energy levels calculated in the jDFT and QC (Quant-Chem) methods, including isosurface images of the QC orbitals for Na$_{20}$, are presented (left). These results but for higher-lying occupied levels for Na$_{20}@$C$_{240}$ are also given (right). }
\label{gpot}
\end{figure*}

\section{Results and Discussions}
\subsection{Ground state structures}
\subsubsection{QC versus jDFT}
On the left part of Figure~\ref{gpot} are displayed the four energy levels ($1s^2 1p^6 1d^{10} 2s^2$) in the harmonic oscillator notation that hold twenty delocalized electrons of Na$_{20}$ calculated in jDFT. These are compared with the energies of ten highest occupied molecular orbitals (HOMO-9 to HOMO) of the cluster obtained in the QC approach. The corresponding 3D isosurface images (shown) of the QC orbitals are found to be of predominantly delocalized ``super-orbital" nature remarkably similar to \textcolor{black}{the jDFT isosurface orbitals displayed in Figure 3.} In particular, the QC orbital shapes also show one nearly spherical $1s$-type, three dumbbell-shaped (three-fold degeneracy lifted) $1p$-type, and five double dumbbell-shaped (from five-fold degeneracy lifted) $1d$-type orbitals. The only exception is the QC analog (HOMO) of the jDFT $2s$ orbital \textcolor{black}{(not shown but is spherical as $1s$)} which exhibits a hybrid character by mixing a $s$-type (spherical blue region) with a $f$-type (yellow region with three-arms) orbital. Note that a spin-up and a spin-down electron occupy each of the QC levels. Comparing the $\na$ energy levels in Fig.\,1, the QC level ordering aligns well with that of jDFT. \textcolor{black}{QC-DFT methods are characteristically known~\cite{vinit17,zhang14} to yield incorrect absolute energies for HOMO while our jDFT HOMO is accurate by adjusting the pseudopotential to match the experimental first ionization potential (FIP) of $\na$~\cite{Chandezon}. Thus,} except for an overall shift of about 0.8 eV, there is a close consistency in the level spacing between the two methods, although QC features narrow splitting due to the broken degeneracy mentioned above. We note that the aforementioned discrepancy in the character of QC HOMO and jDFT $2s$ \textcolor{black}{may have insignificant consequence in the context of Na$_{20}$@C$_{240}$ for the following reason. As we see below, QC supports six outer electrons of Na$_{20}$ to transfer to $\ful$ to equilibrate the ground state of the endofullerene and we calculate Na$_{20}$@C$_{240}$ in jDFT in that equilibrium configuration. Here the Green's function formalism explicitly requires the occupied states and also the continuum states, see \eq{suscep} through \eq{green}}.

The ground state geometry optimization of $\naful$ in the QC predicts that six $\na$ electrons are transferred to the $\ful$ shell. This transfer occurs due to the following reason. The optimized structures of {\em independent} systems indicate that \textcolor{black}{the top three levels} of $\na$ are energetically above the three-fold degenerate (considering the up-down spin pair being a fold) LUMO of $\ful$ accounting for six electrons. Therefore, the composite system relaxes further to its ground state upon a six-electron transfer. Some higher-lying occupied levels of $\naful$ obtained in the simulation, along with the isosurface images of corresponding molecular orbitals, are presented on the right part of Fig.\,1. Note that three outermost levels, HOMO to HOMO-2, are of predominantly delocalized $d$-type orbitals belonging to $\na$, although HOMO-2 is somewhat hybridized with $\ful$. The next three orbitals, HOMO-3 to HOMO-5, housing another six electrons are of $\ful$ character. The following HOMO-6 to HOMO-8, on the other hand, closely present delocalized $p$-type $\na$ orbitals. Subsequently, HOMO-9 to HOMO-22 account for 28 $\ful$ electrons above the $s$-type $\na$ state in HOMO-23. These details of the QC result are utilized to compose the ground state of the system in jDFT to obtain the closest match as described in the following.

The replication of the QC ground state by jDFT entailed the transfer of six $\na$ electrons to $\ful$ leaving fourteen electrons (1$s^2$1$p^6$1$d^{6}$) in the cluster. Hence, the system, while electrically neutral in its entirety, can be truly characterized as Na$_{20}^{+6}$@C$_{240}^{-6}$. This resulted in the otherwise empty $1x$ state of $\ful$ corresponding to $l=19$ in jDFT to energetically move down in the endofullerene to become a partially occupied state; see jDFT energy levels in Fig.\,1. In $\naful$, we denote the $\na$-type states as $nl$@ and the $\ful$-type states as @$nl$. \textcolor{black}{Noticeably, the $1d$@, $1p$@, and $1s$@ jDFT states (see the orbital images in Fig.\,3) closely reproduce QC orbitals in their $\na$ character.} The ordering as well as the spacing of the energy levels between jDFT and QC are fairly comparable (Fig.\,1). Also, the charge transferred state @$1x$ has a binding energy of $\sim$3.97 eV, which is proximate to FIP of $\sim$3.58 eV of the system corresponding to the binding energy of outer $1d$@. In general, the jDFT formalism helps in gaining a better insight into the coupling of $\na$ and $\ful$ that we describe below.
\begin{figure}[ht!]
\includegraphics[width=\linewidth]{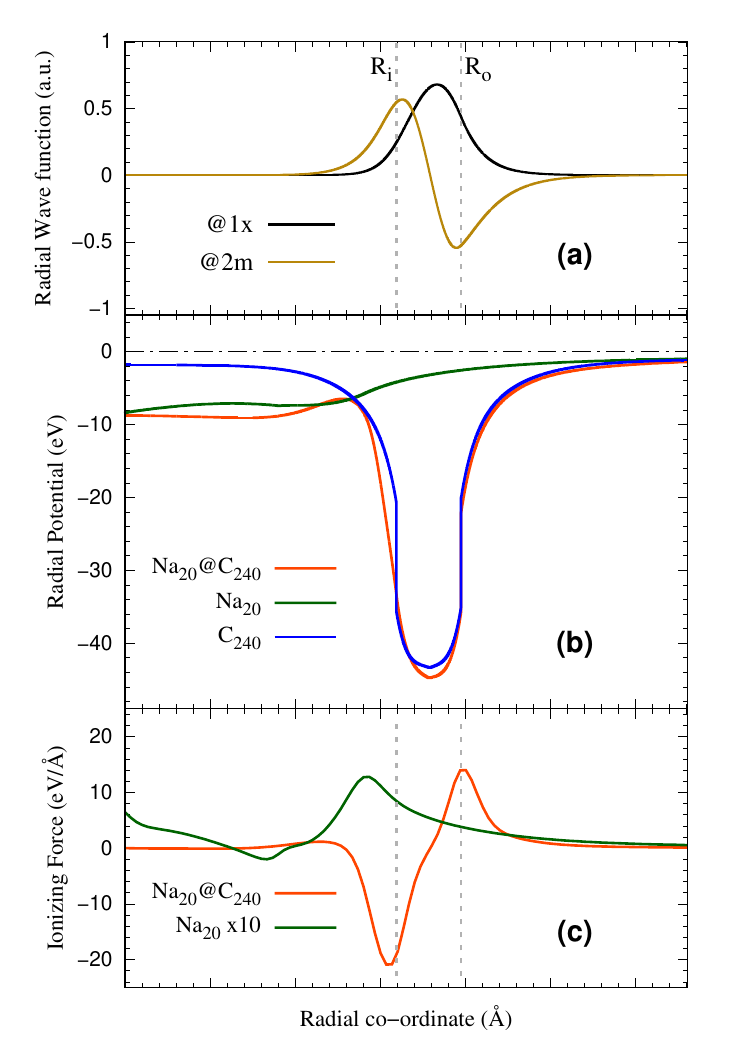}
\caption{Radial wave functions of representative $\sigma$ and $\pi$ occupied levels of $\naful$ calculated in jDFT (a). The jDFT radial potentials obtained for Na$_{20}$, C$_{240}$ and $\naful$ \textcolor{black}{(b)}.\textcolor{black}{The derivative of the radial potential for Na$_{20}$ (multiplied by a factor of 10) and Na$_{20}$@C$_{240}$ (c).}} 
\label{wfpot1}
\end{figure}
\begin{figure}[ht!]
\includegraphics[width=\linewidth]{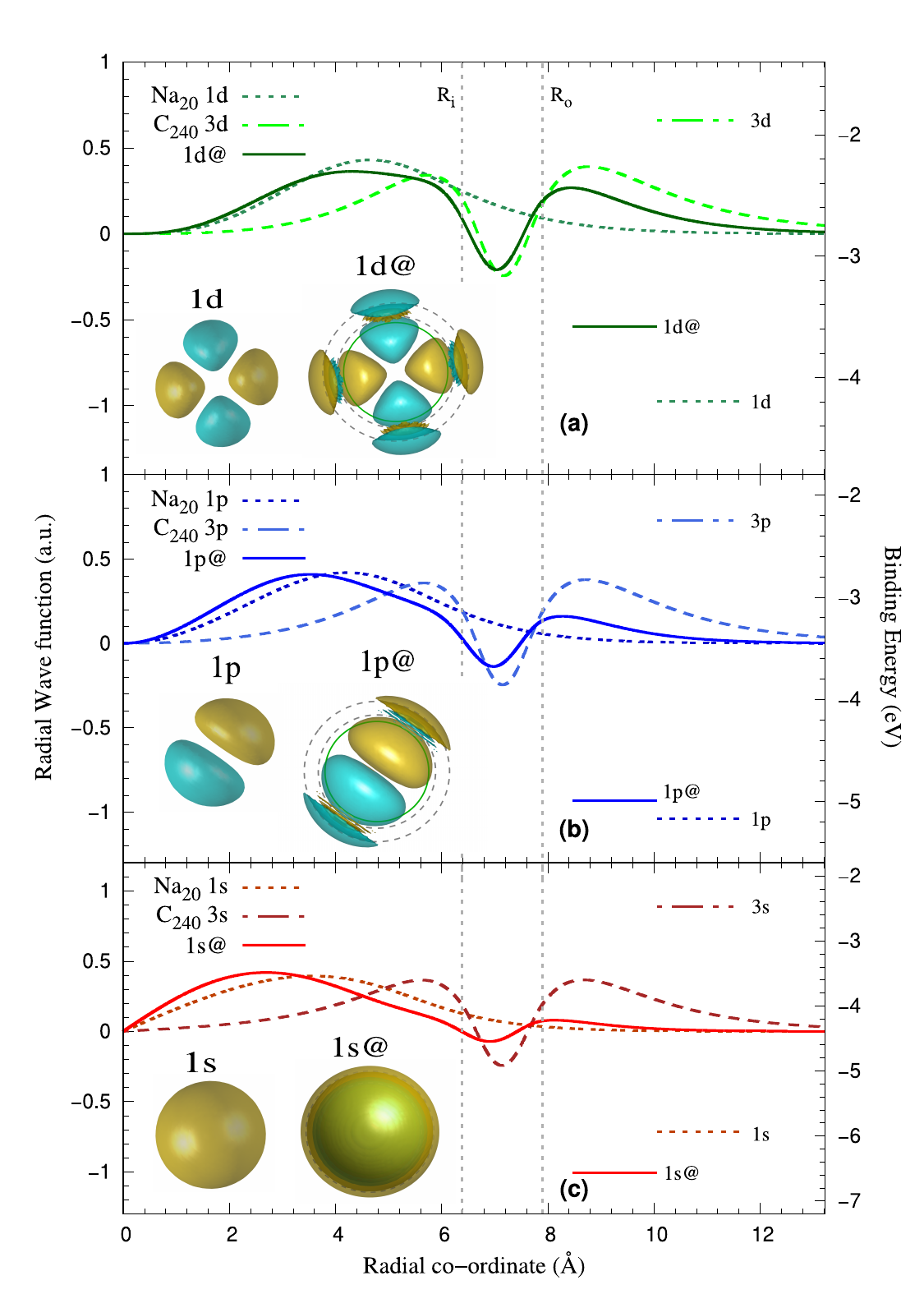}
\caption{Radial wave functions of $s$, $p$ and $d$ angular symmetry calculated in jDFT for pristine Na$_{20}$ and C$_{240}$ (unoccupied) compared with the $\naful$ \textcolor{black}{symmetric} hybrid levels: $l=2$ (a), $l=1$ (b) and $l=0$ (c). \textcolor{black}{The isosurface images of the occupied orbitals are also presented.}} 
\label{wfpot2}
\end{figure}

\subsubsection{Hybridization with SAMO-type orbitals}

Two representative jDFT radial wave functions of Na$_{20}$@C$_{240}$ \textcolor{black}{occupied levels} are presented in Figure \ref{wfpot1}(a). These are $\ful$-type @$1x$ and @$2m$ levels of the shapes of typical $\sigma$ (0-node) and $\pi$ (1-node) character, respectively, \textcolor{black}{of the $\sigma$ and $\pi$ band~\cite{choi2017}}. Their spatial distribution appears within and closely around the inner (R$_{i}$) and outer (R$_{o}$) radius of the $\ful$ shell. In Fig.\,\textcolor{black}{2(b)} are displayed the occupancy-weighted radial averages of \eq{dft-pot} over the subshells. On the other hand, the jDFT radial wave functions of the unoccupied C$_{240}$ $n=3$ levels, $3d$, $3p$ and $3s$, are shown in Fig.\ref{wfpot2}. Unlike the occupied levels in Fig.\,2(a), these $\delta$-type (2-node) empty levels have distinctly different amplitude distribution. They show \textcolor{black}{significant} amplitudes in the shallow potential regions [Fig.\ref{wfpot1}(\textcolor{black}{b})] both inside (i.e. $r<R_{i}$) the hollow cavity and outside the shell. Similar subnanometer-size atom-like orbitals were first detected experimentally in C$_{60}$ on a Cu surface by Petek \textit{et al.}~\cite{Petek2008} which are known as super-atom molecular orbitals (SAMOs). For $n=3$, these resemble the harmonic shapes of $s$, $p$ and $d$ SAMOs obtained using DFT and few other {\it ab\,initio} calculations in a variety of hollow molecules~\cite{Zhao2009} and endohedral anions~\cite{Gromov2015}. 

The transfer of six electrons from $\na$ to $\ful$ in the system somewhat nontrivially reshapes the self-consistent field ground state potential. As seen in Fig.\,2\textcolor{black}{(b)}, the radial potential of $\naful$ is not a mere sum of the potentials of individual systems. Nevertheless, the radial shape of $1d$@, $1p$@ and $1s$@ jDFT wave functions of $\naful$ can still be \textcolor{black}{approximately modeled} by a two-state symmetric (bonding) hydridization of $\na$ $1d$, $1p$ and $1s$ with the SAMO-type $3d$, $3p$ and $3s$ of $\ful$ respectively. \textcolor{black}{In an exact formalism,} this hybridization within same $l$-values is governed by the orthonormality principle. As such, the symmetric cluster-fullerene hybrids can be represented as:  
\begin{equation}
\phi_{nl@}=\sqrt{a}\phi_{1l}^{\scriptsize \na}+\sqrt{1-a}\phi_{3l}^{\scriptsize \ful}
\label{hybrid}
\end{equation}
with $a\in [0,1]$ is the appropriate normalization coefficient for mixing and $l$=0($s$), 1($p$), and 2($d$). The radial wave functions of these participant and resulting hybrid states are presented in Fig.\ref{wfpot2} in individual panels. The best fittings using \eq{hybrid} suggests approximately 70-30\%, 80-20\%, and 90-10\% mixing for $d$, $p$, and $s$ states respectively with dominating $\na$ character. \textcolor{black}{Fig.\ref{wfpot2} further includes isosurface jDFT images of the hybrid orbitals. This mixing of $\na$ states with $\ful$ SAMO-type states will uncover an underlying coherence process in the plasmon RICD mechanism in Subsection III C.} We further note that the antisymmetric (anti-bonding) hybrids are not relevant, since they being unoccupied do not directly participate in the photoionization process. 

The binding energies of these levels for Na$_{20}$, C$_{240}$ and Na$_{20}$@C$_{240}$ are shown by color-coded horizontal lines on each panels of Fig.\ref{wfpot2}. In the spirit of the perturbation theory, the hybridization can be determined by (i) the energy proximity of unperturbed (free system) participant levels and (ii) their wave function overlaps. As seen, the energy separations are large which do not support the hybridization. But owing to the SAMO-character of $\ful$ states, their overlaps with $\na$ wave functions are large, which somewhat favors the mixing. We note that the $1s$@ level is slightly more bound compared to $1s$, whereas $1p$@ and $1d$@ are less bound compared to $1p$ and $1d$, respectively. In general, the transfer of electrons impacts the properties of the levels: $1s$@, being spatially closer to the Na$_{20}^{+6}$ cation, is more localized by the cationic potential. But $1d$@ and $1p$@ being farther from the center and rather weakly bound find significant energy ``room" in the C$_{240}$ region which supports their slightly stronger hybridization. This leads to some reduction of $1p$@ and $1d$@ binding in the combined system. On the other hand, while the transfer hardly affects the C$_{240}$ $\pi$ levels, the $\sigma$ levels show an upward shift due to a somewhat increased electron-electron repulsion on the $\ful$ region. 
\begin{figure}[ht!]
\includegraphics[width=\linewidth]{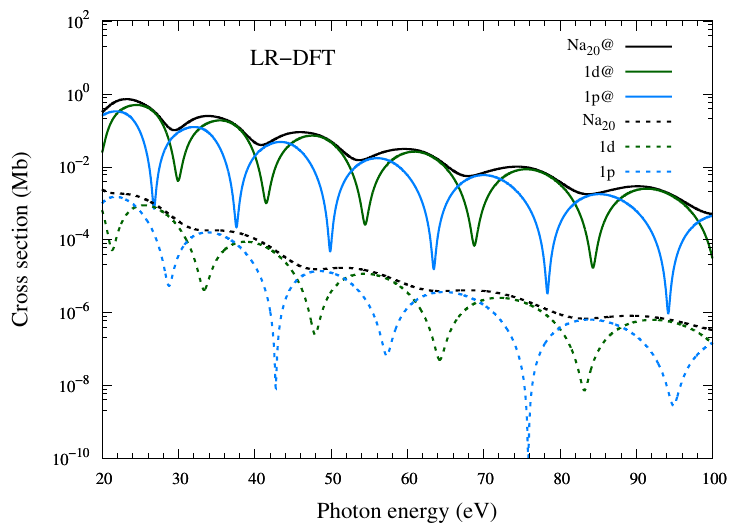}
\caption{LR-DFT photoionization cross section for Na$_{20}$ and Na$_{20}$@ levels. Also shown are the total cross sections of Na$_{20}$ and Na$_{20}$@.} 
\label{lr-dft}
\end{figure}

\subsection{Photoionization in IP Frame}

Figure \ref{lr-dft} shows the LR-DFT $1d$@ and $1p$@ cross sections at high photon energies far beyond the plasmon resonance region where the electron correlation becomes negligible. \textcolor{black} {Note that in the LR-DFT framework, there is no coupling between Na$_{20}$ and C$_{240}$ ionization channels.} The oscillations in these \textcolor{black}{LR-DFT} cross sections can be understood in the IP framework. These oscillations are linked to a diffraction mechanism driven by the $\naful$ geometry, as are generally well known in the photoionization of symmetric clusters~\cite{mccune2008}. 

Within the framework of the acceleration gauge formalism, we can rewrite the LR-DFT radial dipole matrix element in Eq.(\ref{pi}) as, 
\begin{equation}
    \mathcal{D}_{nl \rightarrow kl'} \sim \Big<{kl'}\Big|\frac{dV}{dr}\Big|{nl}\Big>.
    \label{pi-matrix}
\end{equation}
 This matrix element embodies the notion that within the potential $V(r)$ describing Na$_{20}$@C$_{240}$, the outgoing photoelectron experiences the ionizing force ($dV/dr$) to break free from the system. This force field offered by the radial potential of the endohedral complex and free Na$_{20}$ are illustrated in Fig.\ref{wfpot1}\textcolor{black}{(c)}. The former reveals that ($dV/dr$) attains large values at R$_{i}$ and R$_{o}$ of C$_{240}$ offering large ionizing force from these locations. This curve also suggests that the Na$_{20}$ region, on the other hand, imparts practically zero force; see that the curve is almost featureless over the central Na$_{20}$ region. However, whether emissions actually occur at $\ful$ shell edges must depend on, as Eq.(\ref{pi-matrix}) suggests, the bound wave function $\phi_{nl}$ being nonzero there. As seen in Fig.\ref{wfpot2}(a), (b), and (c), the hybrid $1d$@, $1p$@, and $1s$@ wave functions closely pass through a node at R$_{i}$ implying that photoemission from the inner edge almost vanishes. In contrast, these wave functions being non-zero at R$_{o}$ support predominant ionization from this location. Therefore, the resulting diffraction occurs only due to the outer edge of $\naful$. Hence, the cross sections should oscillate in a frequency of 2R$_{o}$, the outer diameter, as a function of the photoelectron momentum. These oscillations (fringes) are seen in Fig.\ref{lr-dft} for $1d$@ and $1p$@ in the photon energy; the similar result for $1s$@ is not shown.
 
 Fig.\ref{lr-dft} further includes the cross sections of the respective levels of free Na$_{20}$ in which the diffraction is known to originate due to the dominant emission from the cluster edge $R_{c}$~\cite{shaik2021}. Since $R_{c} < R_{o}$, the oscillations in $nl$@ is somewhat shorter. However, as can be noted, the non-oscillatory (average) background strength of the cross sections are enhanced by roughly two orders of magnitude in $nl$@ compared to free Na$_{20}$. Referring to the potential gradient in acceleration gauge amplitude (Eq.(\ref{pi-matrix})), this strength is largely drawn from the force around the peak with ``ridges" on either side of the peak. Indeed, the $\naful$ peak height at R$_{o}$ in Fig.\ref{wfpot1}\textcolor{black}{(c)} being about ten times larger than that at $R_{c}$ of Na$_{20}$ justifies the increase by about 10$^{2}$ in the $nl$@ cross sections. Therefore, we stress that this enhancement at the IP level is a direct consequence of \textcolor{black}{(i) the $\na$-$\ful$ coupling induced modification of the potential shape providing much stronger ionizing force and (ii) the wave function hybridization.}
 \begin{figure}[ht!]
\includegraphics[width=\linewidth]{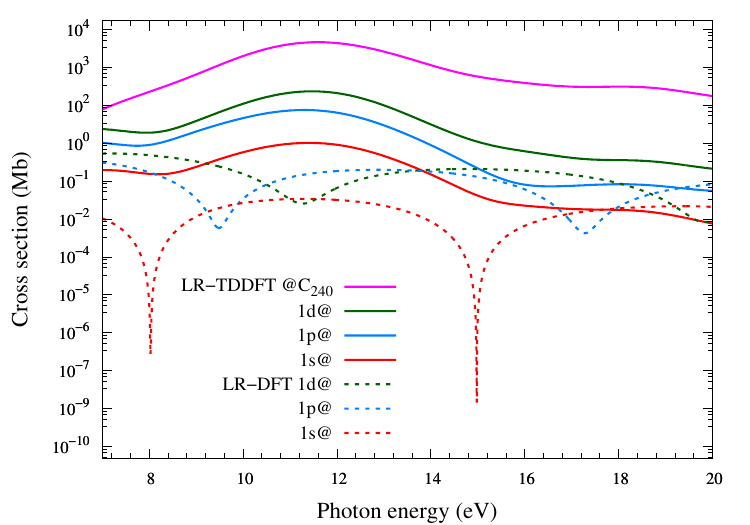}
\caption{Comparison of LR-TDDFT and LR-DFT photoionization cross section of Na$_{20}$@ levels over the lower energy GPR region, along with the LR-TDDFT @C$_{240}$ cross section.} 
\label{picd-gpr}
\end{figure}
\begin{figure}[ht!]
\includegraphics[width=\linewidth]{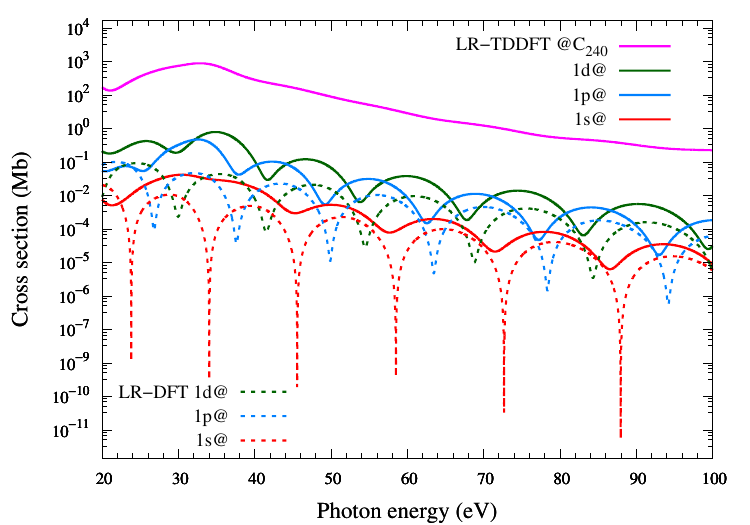}
\caption{Comparison of LR-TDDFT and LR-DFT photoionization cross section of Na$_{20}$@ levels over the higher energy HPR region, along with the LR-TDDFT @C$_{240}$ cross section.} 
\label{picd-hpr}
\end{figure}
 
\subsection{Photoionization by Plasmonic Resonant ICD}
Figure \ref{picd-gpr} shows LR-TDDFT \textcolor{black}{and LR-DFT cross sections for the confined Na$_{20}$ levels over $\ful$ GPR energy region along with the @$C_{240}$ total cross section}. The results are convoluted with a Gaussian function of width 0.4 eV. This was done to mask the narrow single-electron inner-shell excitation Fano-type resonances~\cite{Koskinen1996,rshaik2023} in order to capture the broad plasmonic profiles in the ionization of these levels. As seen, the Na$_{20}$@ levels show spectacular enhancements, peaking around 12 eV, in the cross section compared to the \textcolor{black}{LR-DFT results}. This effect is due to the RICD of the C$_{240}$ GPR through the continuum of the Na$_{20}$@ levels. Incidentally, $\na$'s indigenous plasmon excites at much lower visible part of the spectrum below its ionization threshold \textcolor{black}{and, therefore, does not interfere with the RICD process. Figure \ref{picd-hpr}, subsequently, presents a similar result of the RICD driven enhancements of $\na$@ levels in the energy range of $\ful$ HPR. Note significant enhancements of the Na$_{20}$@ cross sections maximizing around 35 eV.} In any case, this RICD induced enhancement, \textcolor{black}{but only for GPR and in the total cross section} was reported in our previous study~\cite{shaik2023}. The current study not only \textcolor{black}{incorporates new result at HPR, but also probes the RICD boosts in individual Na$_{20}$-like levels for both the resonances. In what follows, we adopt an approach to characterize a fundamental mechanism for deeper understanding of the plasmon RICD.} 

A classic framework to diagnose the RICD enhancement is based on the well-known interchannel coupling approach introduced by Fano~\cite{Fano1961}. \textcolor{black}{We emphasize that we use this framework only to draw insights, while the numerical calculation is performed in LR-TDDFT. Further, while the framework was extensively discussed earlier to probe single-electron excitation RICD~\cite{javani2014,magrakvelidze2016,de2016,de2021}, we simply re-purpose it here for the plasmonic RICD.} We are interested in the  RICD amplitude of the plasmon vacancy decay, after a collective excitation ($0 \rightarrow m$), through a typical Na$_{20}$ ionization channel ($nl @ \rightarrow kl'$). The effect of electron correlation \textcolor{black}{enters in this amplitude numerically via the operator $\delta V$ in Eq.\ (\ref{pi}). However, a far insightful way to embody this correlation in the amplitude is by Fano} interchannel coupling of the plasmon (pl) excitation channel with a continuum (c) channel of Na$_{20}$. Thus, following Ref.\ \cite{shaik2023} , the RICD amplitude $M_{nl@}^{\mbox{\scriptsize pl-c}}(E)$ can be written as
\begin{equation}
    \mathcal{M}_{nl @ \rightarrow kl'}^{\mbox{\scriptsize pl-c}}(E) \sim \frac{\Big<{o \rightarrow m}\Big|\frac{1}{|\vec{r}_{\mbox{\scriptsize pl}}-\vec{r}_{nl@}|}\Big|\psi_{nl @ \rightarrow kl'}(E)\Big>}{E-(E_{m}-E_{0})}\mathcal{D}_{0 \rightarrow m}.
    \label{fano}
\end{equation}
In \eq{fano}, $E_{m}-E_{0}$ and $\mathcal{D}_{0 \rightarrow m}$ are, respectively, the plasmon excitation energy and the LR-TDDFT amplitude of the @$\ful$ (native) plasmon ionization. $E$ is the photon energy corresponding to the $nl @ \rightarrow kl'$ transition and $\psi$ refers to the IP (LR-DFT) channel wave function associated with this transition.  $\vec{r}_{nl@}$ and $\vec{r}_{\mbox{\scriptsize pl}}$ are the co-ordinate vectors of the photoelectron and the plasmon quasiparticle in the decay process, respectively. Clearly, the Coulombic matrix element in the numerator of Eq.(\ref{fano}) mediates the channel interaction. This, therefore, serves as a conduit for transferring a part of plasmonic strength across to decay through $\na$@ continuum channels. The mechanism can also be visualized as a virtual energy relocation from the C$_{240}$ plasmon de-excitation to the Na$_{20}$ ionization process, \textcolor{black}{as in a participant RICD~\cite{Jahnke2020}}. This then effectuates the RICD enhancement in LR-TDDFT $nl$@ subshell cross sections in Figs.\ref{picd-gpr} and \ref{picd-hpr}.

At this stage, let us recall \eq{hybrid} that represents the ground state hybridization of $\na$ levels with SAMO-type $\ful$ levels.  Following this, the hybridization of the final state channel wave function $\psi_{nl@}$ should assume the form,
\begin{equation}
   |\psi_{nl @ \rightarrow kl'}> = \sqrt{a}|\psi_{nl \rightarrow kl'}^{\scriptsize \na}> + \sqrt{1-a}|\psi_{3l \rightarrow kl'}^{\scriptsize \ful}>.
  \label{transition-hybrid}
\end{equation}
Evidently, as in \eq{hybrid}, the Na$_{20}$ portion of the continuum channel still remains dominant given the stronger share of the Na$_{20}$ character in $1l$@ bound states (Fig.\ \ref{wfpot2}). 

Inserting Eq.\ (\ref{transition-hybrid}) in Eq.\ (\ref{fano}), and assuming that the overlap between a pure $\na$ and $\ful$ bound state wave function is rather weak, we may approximately separate the integration regions of $\na$ and $\ful$ to obtain, 
\begin{widetext}
\begin{equation}
 \mathcal{M}_{nl @ \rightarrow kl'}^{\mbox{\scriptsize pl-c}}(E)
 \sim \Bigg[\sqrt{a}\frac{\Big<{0 \rightarrow m}\Big|\frac{1}{|\vec{r}_{\mbox{\scriptsize pl}}-\vec{r}_{nl@}|}\Big|\psi_{nl \rightarrow kl'}^{\scriptsize \na}(E)\Big>}{E-(E_{m}-E_{0})} 
       + \sqrt{1-a}\frac{\Big<{0 \rightarrow m}\Big|\frac{1}{|\vec{r}_{\mbox{\scriptsize pl}}-\vec{r}_{@nl}|}\Big|\psi_{3l \rightarrow kl'}^{\scriptsize \ful}(E)\Big>}{E-(E_{m}-E_{0})}\Bigg]\mathcal{D}_{0 \rightarrow m}.
    \label{fano-picd}
\end{equation}
\end{widetext}

 Eq.(\ref{fano-picd}) can be interpreted in the following way. The incident XUV radiation excites a giant plasmon in C$_{240}$. Of course a dominant strength of this collective hole decays through the fullerene's native ionization channels. This is the regular plasmonic Auger-type decay and are not covered in \eq{fano-picd}. This equation, however, includes two processes: The first term on the right denotes the pure RICD amplitude of the plasmon through $\na$'s $nl\rightarrow kl'$ channels. The second term, on the other hand, embodies the decay via the continuum of $\ful$'s SAMO-type states. However, the former being significantly strong dominates the character of the decay. But the latter, even though far weaker, embodies a novel Auger-type decay route which is induced solely due to the hybridization in the system and can be termed resonant SAMO-type Auger decay (RSAD). For a free $\ful$, this decay mode is forbidden, since in that case the SAMO-type states being unoccupied cannot ionize. Evidently, \eq{fano-picd} creates a {\em quantum coherence} condition for the interference between these two terms at the level of the cross section, that involves the modulus square of \eq{fano-picd}. Thus, the enhancement in the Na$_{20}$@ subshell photoionization is due to the coherence between a pure but dominant RICD and a novel but weak RSAD. This characterization is a unique consequence of the coupling between $\na$ and $\ful$. 

 \textcolor{black}{A remark is in order, however. As we noted, the delocalized shape of $\na$@ hybrid orbitals in the central $\na$ region is similar between jDFT and QC (Fig.1 versus 3). For the amplitudes outside this region, while the jDFT realization is SAMO-type, QC yields an atomistic hybridization (amplitudes clustered on C atoms in $\naful$ best seen for HOMO-2 in Fig.1). However, although manifested differently, the core picture of coupling between the plasmonic channels of $\ful$ and the ionization channels of $\na$ via hybridization, as revealed by \eq{fano-picd}, seems to be the key on both tracks. This suggests that the resonant ICD-Auger coherence is a robust effect.}
 \begin{figure}[ht!]
\includegraphics[width=\linewidth]{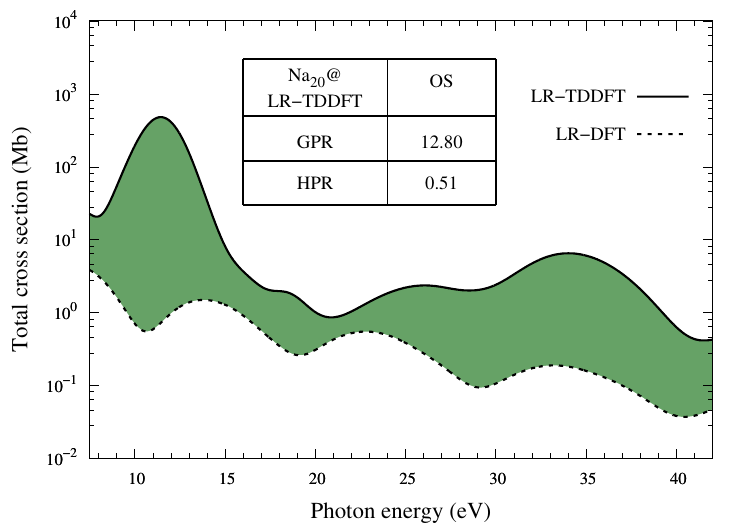}
\caption{Smoothed LR-TDDFT versus LR-DFT cross sections of Na$_{20}$@ over the energies covering both plasmons. The oscillator strengths exhausted by the GPR and HPR of the green shaded region are provided.} 
\label{picd-both}
\end{figure}

\subsection{\textcolor{black}{Plasmon RICD oscillator strength}}

\textcolor{black}{Figure \ref{picd-both} compares the smoothed total $\na$@ cross section LR-TDDFT versus LR-DFT featuring RICD of both GPR and HPR. The comparison delineates the remarkable strength both the plasmons transfer through the RICD channel. The oscillator strength (OS) that the GPR ICD, pivoting around 12 eV, exhausts is 12.8 inducing a structure of relatively narrower bandwidth or longer lifetime. This OS value is large and comparable with the electron number of fourteen associated with the ionic $\na$@. On the other hand, the OS that the RICD harvests from HPR is about 0.51, while spreading over a wider energy range around 35 eV. Even though this value is much smaller, the rise in the cross section is up to an order of magnitude from the LR-DFT estimate. Results in Fig.\ref{picd-both} along with their subshell-differential RICD contributions in Figs.\ref{picd-gpr} and \ref{picd-hpr} may be accessed in photoelectron spectroscopy. Measurements can employ the standard VMI coincidence technique~\cite{basnayake2022} to selectively register photoions or photoelectrons by mass and velocity, as discussed in~\cite{shaik2023}.}  

\section{Conclusions}

 In summary, we perform an {\it ab\,initio} quantum chemical calculation of $\naful$ molecular ground state to establish a benchmark description. A jellium-based DFT model is then constructed to best reproduce the description. A $\na$-$\ful$ hybridization occurs with varieties of consequences. The ground state of the endofullerene reveals that there is a mixing of occupied Na$_{20}$ levels with the unoccupied SAMOs of C$_{240}$. These hybrid states of the Na$_{20}$ relocate some electron probability in the C$_{240}$ shell-region and outside it. This suggests that the occupied electrons of Na$_{20}$ are somewhat drawn towards C$_{240}$. Hybridization-induced redistribution of the ionizing force field within the system brings a remarkable outcome: The average background strength of the (oscillatory) photoionization cross sections of individual Na$_{20}$-type levels are enhanced by roughly two orders of magnitude compared to free Na$_{20}$. This effect is already captured in the independent particle model (LR-DFT), where the electron correlation is omitted. However, when the electron correlation is included via LR-TDDFT, results exhibit spectacular enhancements in the ionization of all $\na$ subshells. These boosts are \textcolor{black}{distributed over the $\na$ subshell photoelectron signals and occur} due to the resonant decay of \textcolor{black}{parts of both the excited giant plasmon and the higher energy plasmon} of $\ful$ . We argue that the amplitude of this process is a sum of a pure, dominant plasmonic RICD through the $\na$ ionization channels and an Auger-type decay via the ionization continuum of some $\ful$ SAMOs, that we termed as RSAD. Even though the latter effect is weak, it only becomes allowed due to the ground state population of these SAMOs driven by hybridization. Consequently, the enhancement in individual $\na$ subshell cross sections incorporate a quantum coherence driven interference between RICD and RSAD, with RICD contribution being dominant. Since electronic-hybridization in plasmonic conjugates is expected to be rather abundant, current results and interpretations may have general relevance in the ICD science.

 \textcolor{black}{It is known that doping metal atoms inside fullerenes leads to stable structures \cite{Wang2024}. However, producing larger fullerenes with higher yields continue to remain a formidable challenge.  A new method involving the isolation of larger fullerenes from fullertube isomers has shown significant promise \cite{Koenig2020,Bourret2023}. In addition, among the various techniques for synthesizing and extracting endofullerenes \cite{Popov2017}, irradiating fullerene films with metal ion beams has proven quite successful \cite{Lee2020}. This method offers the possibility to implant multiple ions, creating clustered encapsulations, by extending the exposure times. Despite the prospect of this method, it might reduce sample yields due to film destruction. We believe that the advancements in technology will improve the current scenario that may lead to designing experiments which can access the plasmon enabled ICD, predicted in this study.}   

\begin{acknowledgments}
The research is supported by the SERB, India, Grant No.\ EMR/2016/002695 (HRV) and by the US National Science Foundation Grant Nos.\ PHY-1806206 (HSC), PHY-2110318 (HSC), and CNS-1624416 (the Bartik HPC Cluster, Northwest Missouri State University). MEM acknowledges the German Research Foundation DFG (FR 2833/79-1) for financial support. FZ acknowledges the support from DFG Research and Training Group via RTG 2247 “Quantum Mechanical Materials Modelling.” MEM. and FZ also thank the computational resources at the HPC Cluster at the University of Oldenburg (Germany)
\end{acknowledgments}

\end{document}